\documentclass{amsart}

\usepackage{amsmath,amsfonts,amssymb}
\usepackage[dvips]{graphicx}

\usepackage{array,delarray}

\usepackage{xspace}
\usepackage{color}
\usepackage{cite}

\usepackage{graphicx}
\usepackage{type1cm}
\usepackage{eso-pic}
\usepackage{color}

\newtheorem{theorem}{Theorem}[section]   % Numbered within each section
\theoremstyle{definition}
\newtheorem{definition}[theorem]{Definition}   % Numbered along with thm

\theoremstyle{remark}
\newtheorem{example}[theorem]{Example}        % Numbered along with thm

\numberwithin{equation}{section}     % Number equations within sections

\usepackage[left=2cm,top=2cm,right=2cm,bottom=2cm]{geometry}

\newcommand\blfootnote[1]{%
  \begingroup
  \renewcommand\thefootnote{}\footnote{#1}%
  \addtocounter{footnote}{-1}%
  \endgroup
}  

\begin{document}
\title[Reduction of AND-NOT Networks]
{Dimension Reduction of Large AND-NOT Network Models}

\author[Veliz-Cuba, Laubenbacher, Aguilar]{Alan~Veliz-Cuba$^\MakeLowercase{ab*}$, Reinhard Laubenbacher$^\MakeLowercase{cd}$, Boris Aguilar$^\MakeLowercase{e}$\\ \\
$^\MakeLowercase{a}$Department of Mathematics, University of Houston
\\
$^\MakeLowercase{b}$Department of Biochemistry and Cell Biology, Rice University
\\
$^\MakeLowercase{c}$Center for Quantitative Medicine, University of Connecticut Health Center\\
$^\MakeLowercase{d}$Jackson Laboratory for Genomic Medicine\\
$^\MakeLowercase{e}$Department of Computer Science,  Virginia Tech
}

\blfootnote{$^*$Corresponding author.
Email: alanavc@math.uh.edu 
}

\begin{abstract}~

\textbf{Objectives}\\
Boolean networks have been used successfully in modeling biological networks and provide a good framework for theoretical analysis. However, the analysis of large networks is not trivial. In order to simplify the analysis of such networks, several model reduction algorithms have been proposed; however, it is not clear if such algorithms scale well with respect to the number of nodes. The goal of this paper is to propose and implement an algorithm for the reduction of AND-NOT network models for the purpose of steady state computation.

\textbf{Methods}\\
Our method of network reduction is the use of ``steady state approximations'' that do not change the number of steady states. Our algorithm is designed to work at the wiring diagram level without the need to evaluate or simplify Boolean functions. Also, our implementation of the algorithm takes advantage of the sparsity typical of discrete models of biological systems.

\textbf{Results}\\
 The main features of our algorithm are that it works at the wiring diagram level, it runs in polynomial time, and it preserves the number of steady states. We used our results to study AND-NOT network models of gene networks and showed that our algorithm greatly simplifies steady state analysis. Furthermore, our algorithm can handle sparse AND-NOT networks with up to 1000000 nodes.

\textbf{Conclusions}\\
The algorithm we propose in this paper allows for fast steady state computation of AND-NOT network models using dimension reduction. Since such networks can arise in qualitative modeling of biological systems, and steady states are important features of mathematical models, it can be a useful tool for model analysis.

\end{abstract}

\maketitle

%%%%%%%%%%%%%%%%%%%%%%%%%%%%%%%%%%%%%
\section{Introduction}
\label{sec-intro}
%%%%%%%%%%%%%%%%%%%%%%%%%%%%%%%%%%%%%
Boolean networks (BN) have been used successfully in modeling biological networks, 
such as gene regulatory networks \cite{AO,Li_cc_2004,Zhang200635,mendozamethod,velizlacop} and provide a good framework for theoretical analysis \cite{Xu20112242,Li20124067}. 
However, the analysis of large networks is not trivial. For example, even the problem of finding or counting steady states has been shown to be hard \cite{Quianchuan,Zhang2007,AkutsuSAT2009}.  Even comprehensive sampling of the phase space is of limited use, once a model
contains 50 or 100 nodes.  

In order to simplify the analysis of such networks, several model reduction algorithms 
have been proposed \cite{redbn,redlm,Saadatpour2010641}. However, it is not clear if such algorithms scale well with respect to the number of nodes. These reduction algorithms are based on using ``steady state approximations'' to remove nodes in a BN. More precisely, to remove a node $i$ in a Boolean network, $f=(f_1,\ldots,f_m):\{0,1\}^n\rightarrow \{0,1\}^n$, one assumes that the $i$-th variable is in steady state and replaces all instances of the $i$-th variable by its Boolean function. For example, we can reduce the BN $f(x_1,x_2,x_3)=(\neg x_3,x_1\vee \neg x_3, x_1\wedge \neg x_2)$, by making the substitution $x_3\rightarrow f_3=x_1\wedge \neg x_2$; then, we obtain the reduced BN $h(x_1,x_2)=(\neg (x_1\wedge \neg x_2),x_1\vee \neg (x_1\wedge \neg x_2))$. This process can be repeated iteratively without changing the number of steady states. 

There are two important aspects in the reduction of BNs. One is the representation of the Boolean functions (e.g. Boolean operators, polynomials, binary decision diagrams, truth tables), and the other is the way in which the reduced network is simplified to ensure that the wiring diagram is consistent with the Boolean functions (e.g. Boolean algebra, polynomial algebra, substitution). It is in these two aspects where algorithms can stop being scalable. For example, although polynomial algebra makes the manipulation of Boolean functions very systematic, the polynomial representation of simple Boolean functions can be large. For instance, storing $x_1\vee x_2\vee\ldots\vee x_k$ and $\neg x_1\wedge\neg x_2\wedge\ldots\wedge\neg x_k$ in polynomial form grows exponentially with respect to $k$. On the other hand, although using Boolean operators can be more intuitive and efficient at representing Boolean functions, their simplification also grows exponentially with respect to the number of variables.

The reduction algorithm in this paper is tailored specifically to the computation of steady states of AND-NOT networks and takes advantage of the sparsity typical of gene regulatory networks. AND-NOT networks are BNs where the functions are of the form $y_{i_1}\wedge y_{i_2} \wedge \cdots \wedge y_{i_r}$ where $y_{i_j}\in \{x_{i_j}, \neg x_{i_j}\}$.
We focus on  AND-NOT networks because they have been shown to be ``general enough'' for modeling and ``simple enough'' for theoretical analysis \cite{bnandnot,SCBN,CBN}.  Also, synthetic AND-NOT gene networks can be designed by coupling synthetic AND gates (e.g. \cite{AND_gates}) and negative regulation. Also, AND-NOT functions are a particular case of nested canalizing functions, which have been proposed as a class of BNs for modeling biological systems \cite{Kauff2,Kauff,Just2004211,Jarrah2007167,Murrugarra2012929}.

Our dimension reduction algorithm for AND-NOT networks has two important properties: First, it preserves all steady state information; more precisely, there is a one-to-one correspondence between the steady states of the original and reduced network. Second, it runs in polynomial time. 

As in previous reduction methods, the main idea of our algorithm is that one can use steady state approximations without changing the number of steady states; however, there are some key differences. First, the only reduction steps that are allowed are those that result in a reduced AND-NOT network. Second, since we are using AND-NOT networks only, we can make additional reductions that cannot be done with other networks. It is important to mention that AND-NOT networks are completely determined by their wiring diagrams. This is important for two reasons: First, we can store AND-NOT networks efficiently using their wiring diagrams and thus avoid the problem that the polynomial representation has. Second, we can state all reduction steps and simplification of the reduced network at the wiring diagram level and thus avoid the problem that the Boolean representation has.

%%%%%%%%%%%%%%%%%%%%%%%%%%%%%%%%%%%%%
\section{Preliminaries}
\label{sec-pre}
%%%%%%%%%%%%%%%%%%%%%%%%%%%%%%%%%%%%%

\subsection{AND-NOT Networks}

\begin{definition}
An \emph{AND-NOT function} is a Boolean function, $b:\{0,1\}^n\rightarrow \{0,1\}$, such that $b$ can be written in the form 
$$
b=b(x_1,\ldots,x_n)=\bigwedge_{j\in P}x_j \wedge \bigwedge_{j\in N}\neg x_j,
$$ 
where $P\cap N=\{\ \}$. If $P=N=\{\ \}$, then $b$ is constant (by convention $\bigwedge_{j\in \{ \ \}} x_j=\bigwedge_{j\in \{ \ \}} \neg x_j=1$). If $i\in P$ ($i\in N$, respectively) we say that $i$ or $x_i$ is a \emph{positive} (\emph{negative}) regulator of $h$ or that it is an activator (repressor). 
An \emph{AND-NOT network} is a BN, $f=(f_1,\ldots,f_n):\{0,1\}^n\rightarrow\{0,1\}^n$, such that $f_i$ is an AND-NOT function or the constant function 0. AND-NOT networks are also called \emph{signed conjunctive networks}.
\end{definition} 

\begin{example}\label{eg:ANBN}
The BN $f:\{0,1\}^6\rightarrow \{0,1\}^6$ given by:\\
$f_1=x_2\wedge x_4\wedge\neg x_5,$\\
$f_2=\neg x_3\wedge \neg x_5 \wedge x_6, $\\
$f_3=0,$\\
$f_4=\neg x_1 \wedge \neg x_5 \wedge x_6,$\\
$f_5=x_6,$\\
$f_6=1,$\\
is an AND-NOT network. For example, $f_1=\bigwedge_{j\in \{2,4\}}x_j \wedge \bigwedge_{j\in \{5\}}\neg x_j$.
\end{example}

\begin{definition}
We say that $x\in\{0,1\}^n$ is a \emph{steady state} or \emph{fixed point} of a BN $f$ if $f(x)=x$; that is, if for all $i=1,\ldots,n$  we have that $f_i(x)=x_i$.
\end{definition}

For example, it is easy to check that 000011 is a steady state of the AND-NOT network in Example \ref{eg:ANBN}.

\begin{definition}
The \emph{extended wiring diagram} of an AND-NOT network is defined as a signed directed graph $G=(V_G,E_G)$ with vertices $V_G=\{0,1,\ldots,n\}$ (or $\{0,x_1,\ldots,x_n\}$) and edges $E_G$ given as follows: $(i,j,+)\in E_G$ ($(i,j,-)\in E_G$, respectively) if $x_i$ is a positive (negative, respectively) regulator of $f_j$. If $f_j=0$, then $(0,j,+)\in E_G$. Positive edges are denoted by ---$\!\blacktriangleright$ and negative edges by ---$\!\bullet$. We will refer to the extend wiring diagram as simply \emph{wiring diagram}.
\end{definition}

\begin{figure}[here]
\centerline{
\framebox{\includegraphics[totalheight=2cm]{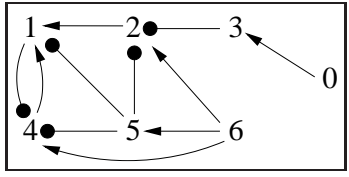}}
}
\caption{Wiring diagram of the AND-NOT network in Example \ref{eg:ANBN}.}
\label{fig:ANBN}
\end{figure}

For example, the wiring diagram of the AND-NOT network in Example \ref{eg:ANBN} is shown in Figure \ref{fig:ANBN}.

%%%%%%%%%%%%%%%%%%%%%%%%%%%%%%%%%%%%%
\section{Reduction of AND-NOT Networks}
\label{sec-red}
%%%%%%%%%%%%%%%%%%%%%%%%%%%%%%%%%%%%%

%%%%%%%%%%%%%%%%%%%%%%%%%%%%%%%%%%%%%
\subsection{Reduction Steps and Algorithm}
\label{sec-red-steps_alg}
%%%%%%%%%%%%%%%%%%%%%%%%%%%%%%%%%%%%%

As mentioned in the Introduction, the idea is to assume that nodes are in steady state and remove them from the network by replacing the variable by the corresponding AND-NOT function. At the wiring diagram level, the idea is to remove nodes and insert edges so that the sign of the edges are ``consistent''. For example, a path $i \textrm{---}\!\!\!\blacktriangleright j \textrm{---}\!\!\bullet k$ should become $i\textrm{---}\!\!\bullet k$ after removing node $j$; and $i\textrm{---}\!\!\bullet j \textrm{---}\!\!\bullet k$ should become $i\textrm{---}\!\!\blacktriangleright  k$ after removing node $j$. The actual rules for doing this depend on the the properties of the node being removed and the incoming and outgoing edges.

\begin{figure}[here]
\centerline{\framebox{\includegraphics[totalheight=8.5cm]{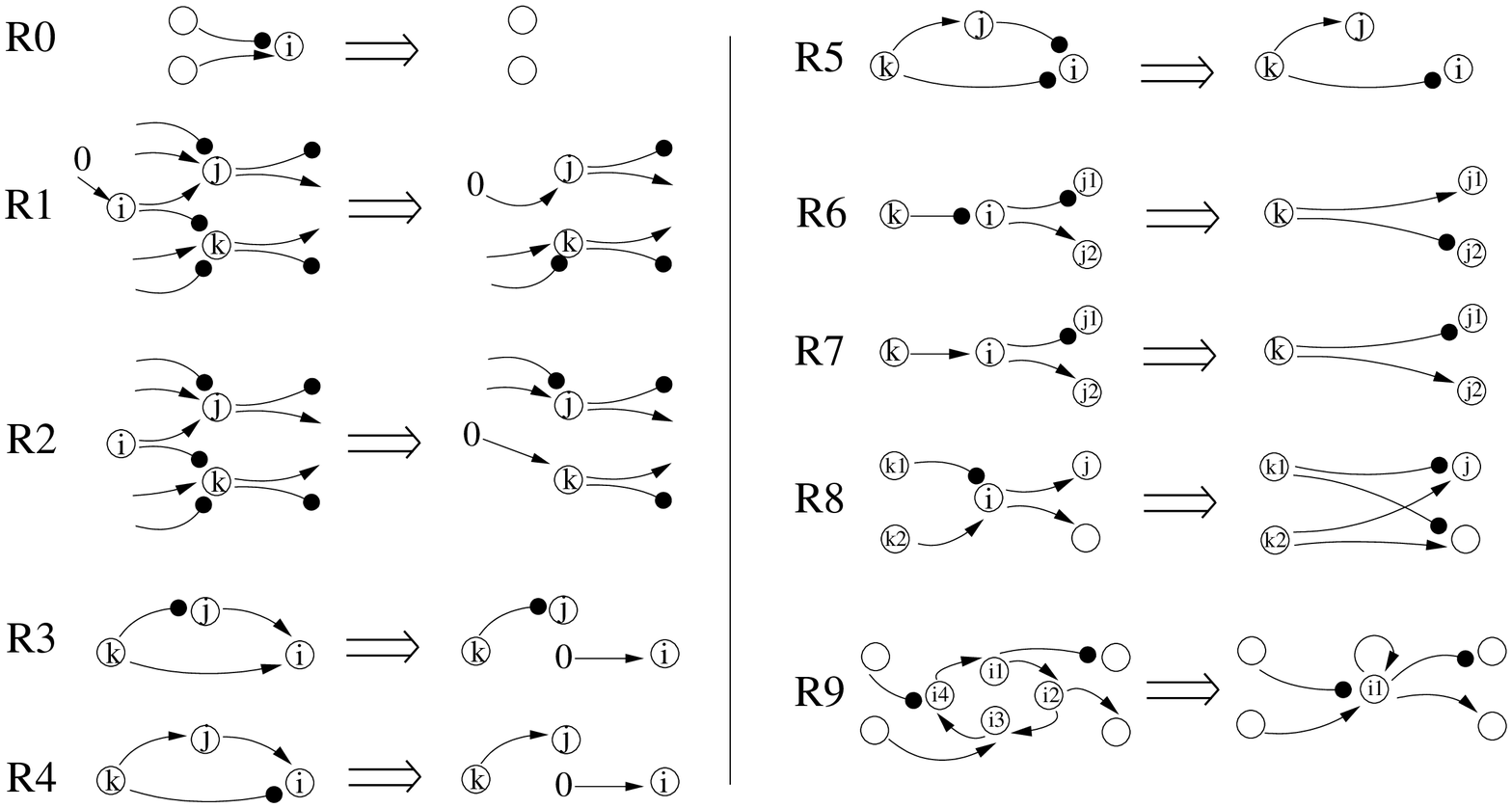}}}
\caption{Reduction steps (before and after). Circles denote nodes. All nodes can have more inputs/outputs not drawn in the figure with the following exceptions: node $i$ in $R0$ does not have any outgoing edges; node $i$ in $R2$ does not have any input; node $i$ in $R5$ does not have any other incoming edge; node $i$ in $R6$ and $R7$ does not have any other incoming edge; node $i$ in $R8$ has positive outgoing edges only.}
\label{fig:algorithm}
\end{figure}

Figure \ref{fig:algorithm} shows the steps at the wiring diagram level. We claim that each of these reduction steps do not change the number of steady states and that the one-to-one correspondence is algorithmic. The proofs follow directly from basic properties of Boolean algebra, so  we only give the idea behind each reduction step.

\begin{itemize}
\item \textbf{Reduction Step $R0$.} Here node $i$ does not have any outgoing edges, so this node does not contribute to the number of steady states and can be removed. Note that given a steady state of the reduced AND-NOT network, the steady state of the original network can be found simply by inserting (in the $i$-th entry) $x_i=f_i$. Note that this reduction step is also valid for general BNs.
\item \textbf{Reduction Step $R1$.} Here we have $f_i=0$; and we remove node $i$ by replacing $x_i$ with $f_i=0$. For example, if $i \textrm{---}\!\!\blacktriangleright j$, then $f_j=x_i\wedge w_j$ for some AND-NOT function $w_j$. By replacing $x_i$ with 0 we obtain $f_j=0\wedge w_j=0$; that is, we add the edge $0\textrm{---}\!\!\blacktriangleright j$ and  remove all other incoming edges of $j$. On the other hand, if $i \textrm{---}\!\!\bullet k$, then $f_k=\neg x_i\wedge w_k$ for some AND-NOT function $w_k$. By replacing $x_i$ with 0 we obtain $f_j=\neg 0\wedge w_k=w_k$; that is, the edge $i \textrm{---}\!\!\bullet k$ is removed and all other edges towards $k$ remain present. Note that given a steady state of the reduced AND-NOT network, the steady state of the original network can be found simply by inserting (in the $i$-th entry) $x_i=0$. We also notice that this reduction step is also valid for general BNs, but not at the wiring diagram level (the wiring diagram of the reduced network depends on the actual Boolean functions). 
\item \textbf{Reduction Step $R2$.} Here we have $f_i=1$; and we remove node $i$ by replacing $x_i$ with $f_i=1$. For example, if $i \textrm{---}\!\!\blacktriangleright j$, then $f_j=x_i\wedge w_j$ for some AND-NOT function $w_j$. By replacing $x_i$ with 1 we obtain $f_j=1\wedge w_j=w_j$; that is, the edge $i \textrm{---}\!\!\bullet j$ is removed and all other edges towards $j$ remain present. On the other hand, if $i \textrm{---}\!\!\bullet k$, then $f_k=\neg x_i\wedge w_k$ for some AND-NOT function $w_k$. By replacing $x_i$ with 1 we obtain $f_j=\neg 1\wedge w_k=0$; that is, we add the edge $0\textrm{---}\!\!\blacktriangleright k$ and  remove all other incoming edges of $k$. Note that given a steady state of the reduced AND-NOT network, the steady state of the original network can be found simply by inserting $x_i=1$. We also notice that this reduction step is also valid for general BNs, but not at the wiring diagram level.
\item \textbf{Reduction Step $R3, R4$.} For $R3$ we have  $f_i=x_j\wedge x_k\wedge w_i$ for some AND-NOT function $w_i$, and a node $j$ with Boolean function $f_j=\neg x_k\wedge w_j$ for some AND-NOT function $w_j$. If we are at a steady state, then we have two cases, either $x_k=0$ or $x_k=1$. If $x_k=0$, then $x_i=f_i=x_j\wedge 0\wedge w_i=0$. If $x_k=1$, then $x_j=f_j=\neg 1\wedge w_j=0$, and then $x_i=0\wedge x_k\wedge w_i=0$. In either case $x_i=0$, so by assuming that $f_i=0$ we are not changing the steady states of the AND-NOT network. That is, we add the edge $0\textrm{---}\!\!\blacktriangleright i$ and remove all other incoming edges of $i$. The reduction step $R4$ is analogous. It is important to mention that this reduction step is not valid for general BNs.
\item \textbf{Reduction Step $R5$.} Here we have that $f_i=\neg x_j \wedge \neg x_k$ and $f_j=x_k\wedge w_j$ for some AND-NOT function $w_j$. If we are at a steady state then we have two cases, either $x_k=0$ or $x_k=1$. If $x_k=0$, then $x_j=f_j=0\wedge w_j=0$ and $x_i=f_i=\neg 0 \wedge \neg 0=1$. If $x_k=1$, then $x_i=f_i=\neg x_j \wedge \neg 1=0$. In either case we have $x_i=\neg x_k$, so by assuming that $f_i=\neg x_k$ we are not changing the steady states.  It is important to mention that this reduction step is not valid for general BNs.
\item \textbf{Reduction Step $R6, R7$.} For $R6$ we have that $f_i=\neg x_k$; and we remove node $i$ by replacing $x_i$ with $f_i=\neg x_k$. For example, if $f_{j_1}=\neg x_i \wedge w_{j_1}$ for some AND-NOT function $w_{j_1}$, then we obtain $f_{j_1}=\neg \neg x_k \wedge w_{j_1}=x_k \wedge w_{j_1}$, which is an AND-NOT function. If $f_{j_2}=x_i \wedge w_{j_2}$ for some AND-NOT function $w_{j_2}$, then we obtain $f_{j_2}= \neg x_k \wedge w_{j_2}$, which is an AND-NOT function as well. Note that given a steady state of the reduced AND-NOT network, the steady state of the original network can be found simply by inserting  $x_i=\neg x_k$. The reduction step $R7$ is analogous.  We notice that this reduction step is also valid for general BNs, but not at the wiring diagram level. Also, the reduction is no longer valid if $i$ has more incoming edges (the reduced network would not be an AND-NOT network).
\item \textbf{Reduction Step $R8$.} Here we have that all outgoing edges of $i$ are positive. and we remove node $i$ by replacing $x_i$ with $f_i$. For example, if $f_i=\neg x_{k_1}\wedge x_{k_2}$ and $f_j=x_i\wedge w_j$ for some AND-NOT function $w_j$, then we obtain $f_j=\neg x_{k_1}\wedge x_{k_2}\wedge w_j$, which is an AND-NOT function. Note that given a steady state of the reduced AND-NOT network, the steady state of the original network can be found simply by inserting $x_i=\neg x_{k_1}\wedge x_{k_2}$. We also notice that this reduction step is also valid for general BNs, but not at the wiring diagram level. It is important to mention that the reduction is no longer valid if $i$ has any negative outgoing edge.
\item \textbf{Reduction Step $R9$.} Here we have a circuit with positive edges only. If we are at a steady state, we have two cases, either $x_{i_1}=0$ or $x_{i_1}=1$. If $x_{i_1}=0$, then it follows that $x_{i_2}=0$ and working forward we obtain that $x_{i_1}=x_{i_2}=\ldots=0$. Similarly, if $x_{i_1}=1$, we obtain that $x_{i_1}=x_{i_2}=\ldots=1$. Thus, by collapsing this circuit into a single node we do not change the number of steady states. Note that given a steady state of the reduced AND-NOT network, the steady state of the original network can be found simply by inserting $(x_{i_2},x_{i_3},x_{i_4},\ldots)=(x_{i_1},x_{i_1},x_{i_1},\ldots)$. Note that this reduction step is no longer valid if one of the edges in the circuit is negative. This reduction step is also valid for general BNs (removing one node at at time), but not at the wiring diagram level.
\end{itemize}

It is important to mention that reduction steps $R0-R8$ cover the possible reductions where we only need to look at incoming and outgoing edges of a node $i$. Other reduction steps could be considered by looking upstream and downstream of a node; for example, one can generalize $R3$ and $R4$ to include longer feedforward loops (e.g. $i_1\textrm{---}\!\!\blacktriangleright i_2\textrm{---}\!\!\blacktriangleright \ldots\textrm{---}\!\!\blacktriangleright i_r$, $i_1 \textrm{---}\!\!\bullet i_r$). However, their detection becomes  computationally expensive. Reduction step $R9$ is included because such circuits can be detected in linear time \cite{Tarjan1972}.

The actual algorithm is given below. The idea is to iteratively apply the reduction steps until the network is no longer reducible (every time a reduction step is used, new reducible nodes can appear). Note that there are many orders in which one can apply the reduction steps, and in some cases they can result in different reduced networks (with the same number of states). Based on the performance of preliminary simulations, the order given below was chosen.

\textbf{Algorithm.} \\
Input: AND-NOT network $G$.\\
Output: List of steady states.
\begin{enumerate}
\item \label{alg:terminal} Use $R0$ to remove terminal nodes.
\item Let $Z=\{j: (0,j,+)\in E_G \textrm{ or } I_j(G)=\{\}\}$. If $Z=\{\}$, then go to \eqref{alg:new_zeros}.
\item Use $R1, R2$ to remove from $G$ the nodes in $Z$.
\item Go to \eqref{alg:terminal}.
\item\label{alg:new_zeros} Use $R3, R4$ to find new nodes with input 0. 
\item If nodes were found in previous step, then go to \eqref{alg:terminal}.
\item Use $R5$ to remove edges.
\item If there are nodes with a single incoming edge only, then use $R6, R7$ to remove them and go to \eqref{alg:terminal}.
\item Find nodes with positive outgoing edges only.
\item \label{alg:end_basic} If nodes were found in previous step, then use $R8$ to remove them and go to \eqref{alg:terminal}.
\item \label{alg:ssc} Find circuits of length greater than 1 with positive edges only. Only use this step once.
\item \label{alg:end_red}If circuits were found in previous step, then reduce them using $R9$ and go to \eqref{alg:terminal}. 
\item\label{alg:ss} Compute the steady states of the reduced AND-NOT network.
\item\label{alg:bs} Use the bijections given by the reduction steps to find the steady states of the original system.
\end{enumerate}

The algorithm has 3 main parts. In (1)-\eqref{alg:end_red} we reduce the AND-NOT network; in  \eqref{alg:ss} we compute the steady states of the reduced AND-NOT network; and in  \eqref{alg:bs} we use these steady states to find the steady states of the initial AND-NOT network. Note that step $R9$ is used only once in the algorithm because none of the other steps create extra circuits.

%%%%%%%%%%%%%%%%%%%%%%%%%%%%%%%%%%%%%
\subsection{Implementation and Computational Complexity}
\label{sec-red-imp}
%%%%%%%%%%%%%%%%%%%%%%%%%%%%%%%%%%%%%

We preliminarily implemented our algorithm in C++ and used the Boost Graph Library to manipulate graphs (code available upon request). We stored the one-to-one correspondence as an acyclic graph so that once the steady states of the reduced network are computed, one simply uses backward substitution to recover the steady states of the original network. The steady states of the reduced AND-NOT network are computed by exhaustive search.

\begin{example}\label{eg:ANBN2} Consider the AND-NOT network given by:\\
$f_1=x_4,$\\
$f_2=x_1\wedge \neg x_3 \wedge x_4,$\\
$f_3=0,$\\
$f_4=x_1$\\
$f_5=\neg x_2 \wedge x_4\wedge x_6,$\\
$f_6=1.$
\end{example}
%
%1
%1
%4  4  1
%ZERO_NODES
%3
%ACYCLIC_GRAPH
%2 5 -1
%4 5 1
%6 5 1
%4 1 1
%4 2 1

The wiring diagrams of this AND-NOT network, the reduced AND-NOT network, and the acyclic graph are in Figure \ref{fig:ANBN2} (see the Appendix for details about the format we use in our implementation). The reduced network is $h(x_4)=x_4$, from which we easily obtain the steady states $x_4=0, 1$. The acyclic graph encodes the following substitution:\\
$x_3=0,$\\
$x_6=1,$\\
$x_1=x_4,$\\
$x_2=x_4,$\\
$x_5=\neg x_2 \wedge x_4\wedge x_6.$

For $x_4=0$ we obtain $x_3=0$, $x_6=1$, $x_2=x_4=0$, $x_1=x_4=0$, $x_5=x_1\wedge\neg x_2\wedge x_6=0\wedge\neg 0\wedge 1=0$; that is, $x=000001$.
For $x_4=1$ we obtain $x_3=0$, $x_6=1$, $x_2=x_4=1$, $x_1=x_4=1$, $x_5=x_1\wedge\neg x_2\wedge x_6=1\wedge\neg 1\wedge 1=0$; that is, $x=110101$.

\begin{figure}[here]
\centerline{
\framebox{\includegraphics[totalheight=2.5cm]{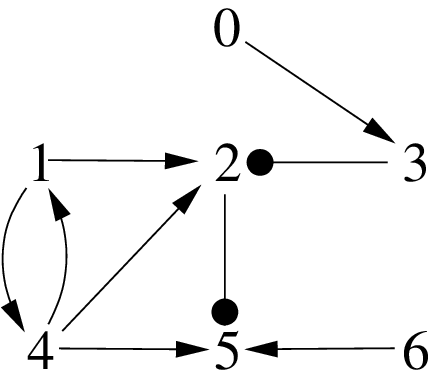}} \framebox{\includegraphics[totalheight=2.5cm]{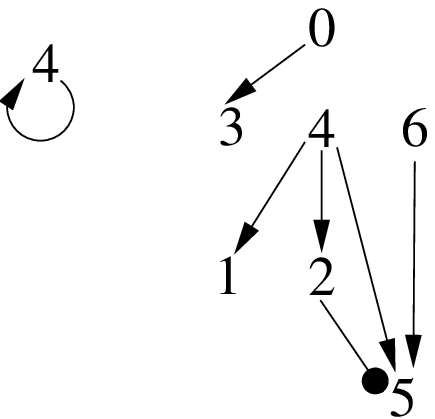}}
}
\caption{Left: Wiring diagram of the AND-NOT network in Example \ref{eg:ANBN2}. Right: The reduced network (with the corresponding acyclic graph encoding the bijection) given by steps (1)-\eqref{alg:end_red}.}
\label{fig:ANBN2}
\end{figure}

Since it is not known the average number of times each pattern in Figure \ref{fig:algorithm} appears in a random AND-NOT network, it is difficult to predict the exact computational complexity of our algorithm. However, we present a heuristic estimation as follows. Let $n$ be the number of nodes and $e$ the number of edges; we denote with $T$ the computational complexity. We first focus on steps (1)-\eqref{alg:end_red} of the algorithm. In the worst case scenario steps \eqref{alg:ssc} and \eqref{alg:end_red} will have to be done at the beginning, this contributes $O(n+e)$ to $T$ \cite{Tarjan1972}. Each detection of an individual pattern and reduction step from (1) to \eqref{alg:end_basic} takes constant time for each node; then, each one of these steps of the algorithm (not counting the ``go to'' statements) contributes $O(n)$ to $T$; and, since we have to repeat this at most $n$ times (counting the ``go to'' statements), we obtain that steps (1)-\eqref{alg:end_basic} contribute $O(n^2)$ to $T$. Thus, steps (1)-\eqref{alg:end_red} contribute $O(n+e+n^2)=O(n^2)$ to $T$. Assuming that one can check in constant time if a state in the reduced network is a steady state, step \eqref{alg:ss} contributes $O(2^m)$ where $m$ is the size of the reduced AND-NOT network. Step \eqref{alg:bs} contributes $O(n^2)$ to $T$.  

Note that the reduction part and backward substitution part contribute $O(n^2)$ to $T$; that is, they run in polynomial time. 

Although $T=O(n^2+2^m)$ is of little improvement if $m\approx n$, Boolean models of biological systems are not arbitrary and have especial properties. For example, they are sparse and have motifs such as feedforward loops (which we considered in $R3-R5$); also, they have few steady states when compared to random networks. Hence, one can argue that for Boolean models of biological systems, the reduced network are likely to be very small. Indeed, for the three Boolean models that we consider in the next section, the size of the reduced networks, $m$, did not exceed $\ln(n)$. The fact that $m$ did not exceed $\ln(n)$ is important, because for family of networks where $m=O(\ln(n))$, we obtain that $T=O(n^2+n^k)$ for some $k>0$. Thus, one can conjecture that under some conditions our algorithm (including steady state computation) runs in polynomial time, but a formal statement and proof of this conjecture is outside the scope of this manuscript. However, our statistical analysis in Section \ref{section:random} supports this.

%%%%%%%%%%%%%%%%%%%%%%%%%%%%%%%%%%%%%
\section{Applications}
\label{sec-application}
%%%%%%%%%%%%%%%%%%%%%%%%%%%%%%%%%%%%%

In this section we apply our reduction algorithm to three published networks and random networks, and demonstrate that it can result in a significant reduction of the network's dimension.  We denote two negative (positive) edges between $i$ and $j$ by a bidirectional negative (positive) edge, $\bullet\!$---$\!\bullet $ ($\blacktriangleleft\!$---$\!\blacktriangleright$); if the edges have different signs we denote them by $\bullet\!$---$\!\blacktriangleright$.

%%%%%%%%%%%%%%%%%
\subsection{Th-lymphocyte Differentiation}\label{section:th}
%%%%%%%%%%%%%%%%%%

\begin{figure}[here]
\centerline{ \framebox{
\includegraphics[totalheight=5cm]{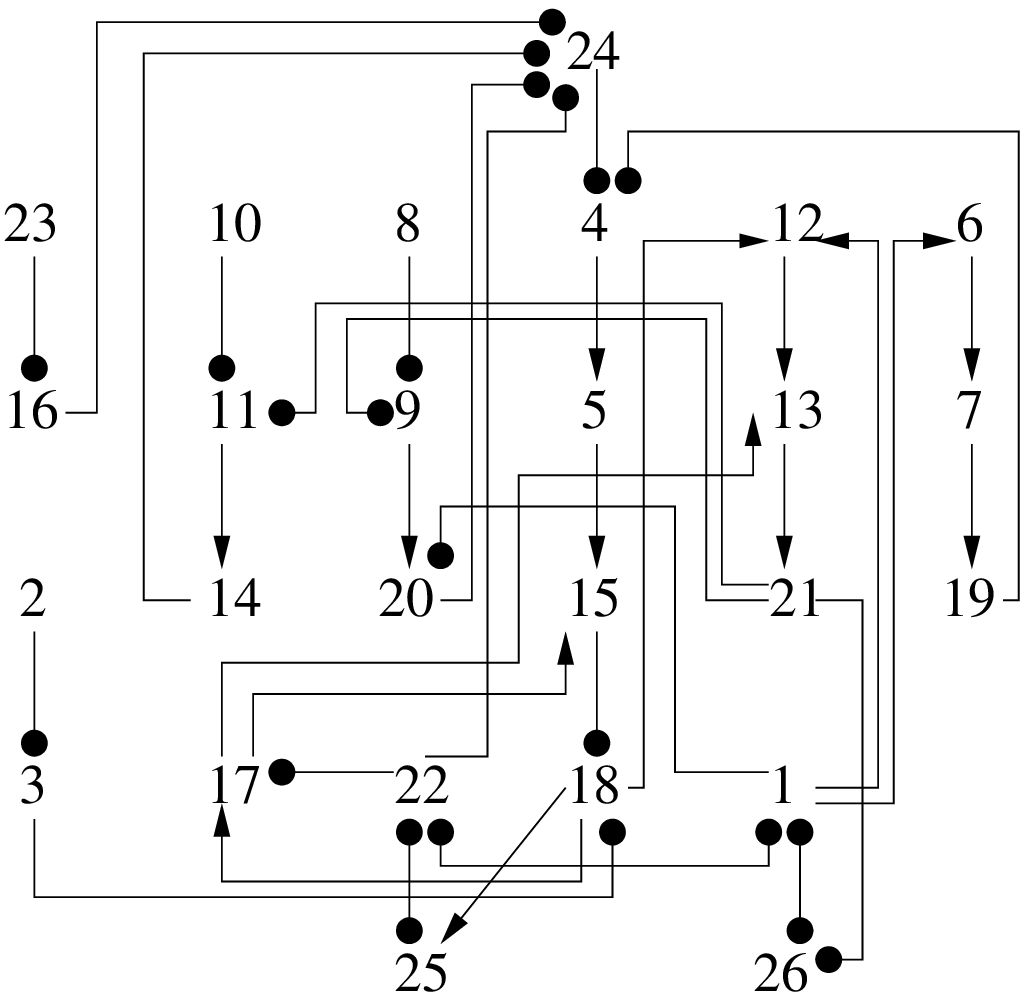} 
}  
 \framebox{
\includegraphics[totalheight=.4cm]{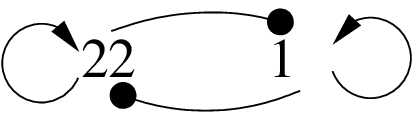}
} }
\caption{AND-NOT model of Th-cell differentiation (left) and the reduced AND-NOT network (right).} \label{fig:Th}
\end{figure}

Here we consider an AND-NOT  model for Th-cell differentiation \cite{mendozamethod,bnandnot}, $f:\{0,1\}^{26}\rightarrow\{0,1\}^{26}$. The wiring diagram is shown in Figure \ref{fig:Th} (left). The state space of this model has $2^{26}\approx 6.7\times 10^7$ states.

$\begin{array}{lllllll}
f_1	 & =& \neg x_{22} \wedge \neg x_{26} &, &
f_2	 & =& 1 \\
f_3   & =& \neg x_2 &, &
f_4	 & =&  \neg x_{19} \wedge \neg x_{24}\\
f_5  & =&  x_4 &, &
f_6  & =&  x_1\\
f_7  & =& x_6  &, &
f_8  & =& 1 \\
f_9  & =& \neg x_8\wedge \neg x_{21} &, &
f_{10}   & =& 1\\
f_{11}   & =& \neg x_{10}\wedge \neg x_{21} &, &
f_{12}   & =& x_1\wedge  x_{18}\\
f_{13} & =& x_{12}\wedge x_{17} &, &
f_{14}  & =& x_{11}\\
f_{15}   & =&  x_5\wedge x_{17} &, &
f_{16} & =&  \neg x_{23}\\
f_{17}  & =& x_{18}\wedge \neg x_{22} &, &
f_{18}  & =& \neg x_3\wedge \neg x_{15}\\
f_{19}  & =& x_7 &, &
f_{20}  & =& \neg x_1\wedge  x_9\\
 f_{21}   & =& x_{13} &, &
 f_{22}   & =&\neg x_1 \wedge \neg x_{25} \\
f_{23}   & =&   1  &, &
f_{24}   & =&  \neg x_{14}\wedge \neg x_{16} \wedge \neg x_{20} \wedge \neg x_{22}  \\
f_{25}   & =&  x_{18} \wedge \neg x_{22}  &, &
f_{26}   & =& \neg x_1 \wedge \neg x_{21}  
\end{array}$

By using our algorithm we reduce this AND-NOT network to the AND-NOT network shown in  Figure \ref{fig:Th} (right), $h:\{0,1\}^{2}\rightarrow\{0,1\}^{2}$, given by $h(x_1,x_{22})=(x_1\wedge \neg x_{22},\neg x_1\wedge x_{22})$. Notice that its state space has only 4 states, which is about 7 orders of magnitude smaller than the original state space. Since this is a small network, it is easy to find its steady states: 00, 01, and 10. Therefore, our results guarantee that the original AND-NOT model has 3 steady states which can be recovered from the steady states of the reduced network. The timing of our implementation was $.00273562s$ (average of 1000000 repetitions).

%%%%%%%%%%%%%%%%%
\subsection{ERBB2 Activation}\label{section:erbb2}
\begin{figure}[here]
\centerline{  \framebox{ 
\includegraphics[totalheight=4.3cm]{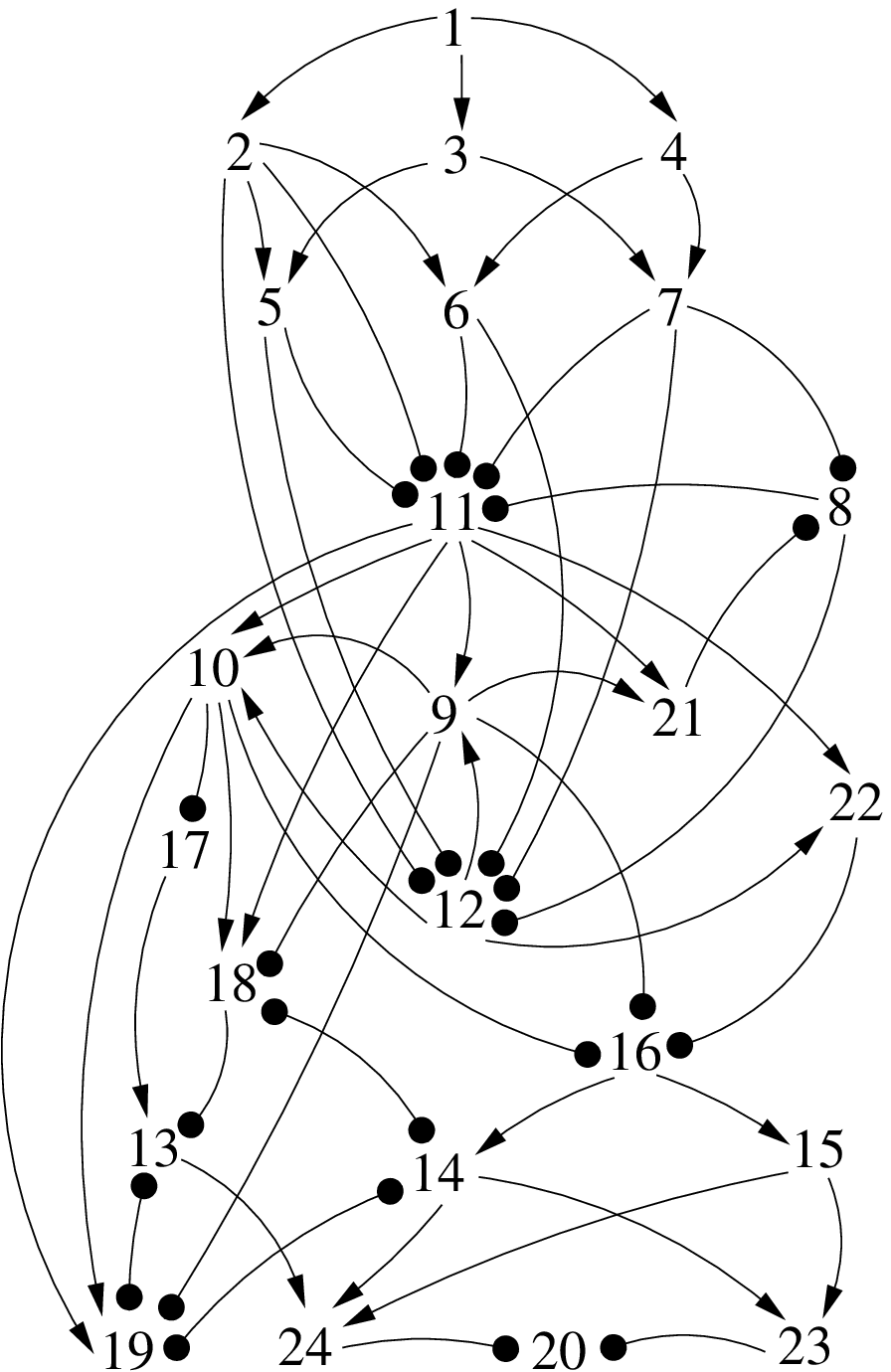}  
} \framebox{ 
\includegraphics[totalheight=.4cm]{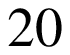}  
}
}
\caption{AND-NOT model of ERBB2 activation (left) and the reduced AND-NOT network (right).} \label{fig:erbb2}
\end{figure}

%%%%%%%%%%%%%%%%%%
Here we consider an AND-NOT network model of ERBB2 activation based on the Boolean model in \cite{erbb2} (left).  The wiring diagram of
the equivalent AND-NOT model is shown in Figure \ref{fig:erbb2}. The state space of this model has $2^{24}\approx 1.6\times 10^7$ states.

$\begin{array}{lllllll}
f_1  &=& 1  &, &
f_2  &=& x_{1} \\
f_3  &=& x_{1}  &, &
f_4  &=& x_{1} \\
f_5  &=& x_{2}  \wedge  x_{3}  &, &
f_6  &=& x_{2}  \wedge  x_{4} \\
f_7  &=& x_{3}  \wedge  x_{4}  &, &
f_8  &=& \neg x_{7} \wedge \neg x_{21} \\
f_9  &=& x_{11}  \wedge  x_{12}  &, &
f_{10} &=& x_{9 } \wedge x_{11}  \wedge  x_{12} \\
f_{11} &=& \neg x_{2}  \wedge  \neg x_{5}  \wedge  \neg x_{6}  \wedge  \neg x_{7}  \wedge  \neg x_{8}  &, &
f_{12} &=& \neg x_{2}  \wedge \neg x_{5} \wedge  \neg x_{6} \wedge  \neg x_{7} \wedge  \neg x_{8} \\
f_{13} &=& x_{17}  \wedge  \neg x_{18}  \wedge  \neg x_{19}  &, &
f_{14} &=& x_{16}  \wedge  \neg x_{18}  \wedge  \neg x_{19} \\
f_{15} &=& x_{16}  &, &
f_{16} &=& \neg x_{9}  \wedge \neg x_{10}  \wedge  \neg x_{22} \\
f_{17} &=& \neg x_{10}  &, &
f_{18} &=& \neg x_{9}  \wedge  x_{10}   \wedge  x_{11} \wedge  \neg x_{14} \\
f_{19} &=& \neg x_{9} \wedge  x_{10} \wedge  x_{11}  \wedge  \neg x_{13}   \wedge  \neg x_{14}  &, &
f_{20} &=& \neg x_{23} \wedge  \neg x_{24} \\
f_{21} &=& x_{9}  \wedge x_{11}  &, &
f_{22} &=& x_{11}  \wedge x_{12} \\
f_{23} &=& x_{14} \wedge  x_{15}  &, &
f_{24} &=& x_{13} \wedge  x_{14}  \wedge  x_{15} 
\end{array}$

By using our algorithm we reduce this AND-NOT network to the AND-NOT network shown in  Figure \ref{fig:erbb2} (right), $h:\{0,1\}^{1}\rightarrow\{0,1\}^{1}$ given by $h(x_{20})=1$.  Notice that its state space has only 2 states, which is about 7 orders of magnitude smaller than the original state space.  It is easy to see that this network has a unique steady state ($x_{20}=1$). 
Therefore, the original AND-NOT model also has a unique steady state, which can be recovered from the steady state of the reduced network.  The timing of the reduction is $.00266462s$  (average of 1000000 repetitions).

%%%%%%%%%%%%%%%%%
\subsection{T-cell receptor}\label{section:thcellrecep}
\begin{figure}[here]
\centerline{  \framebox{ 
\includegraphics[totalheight=5cm]{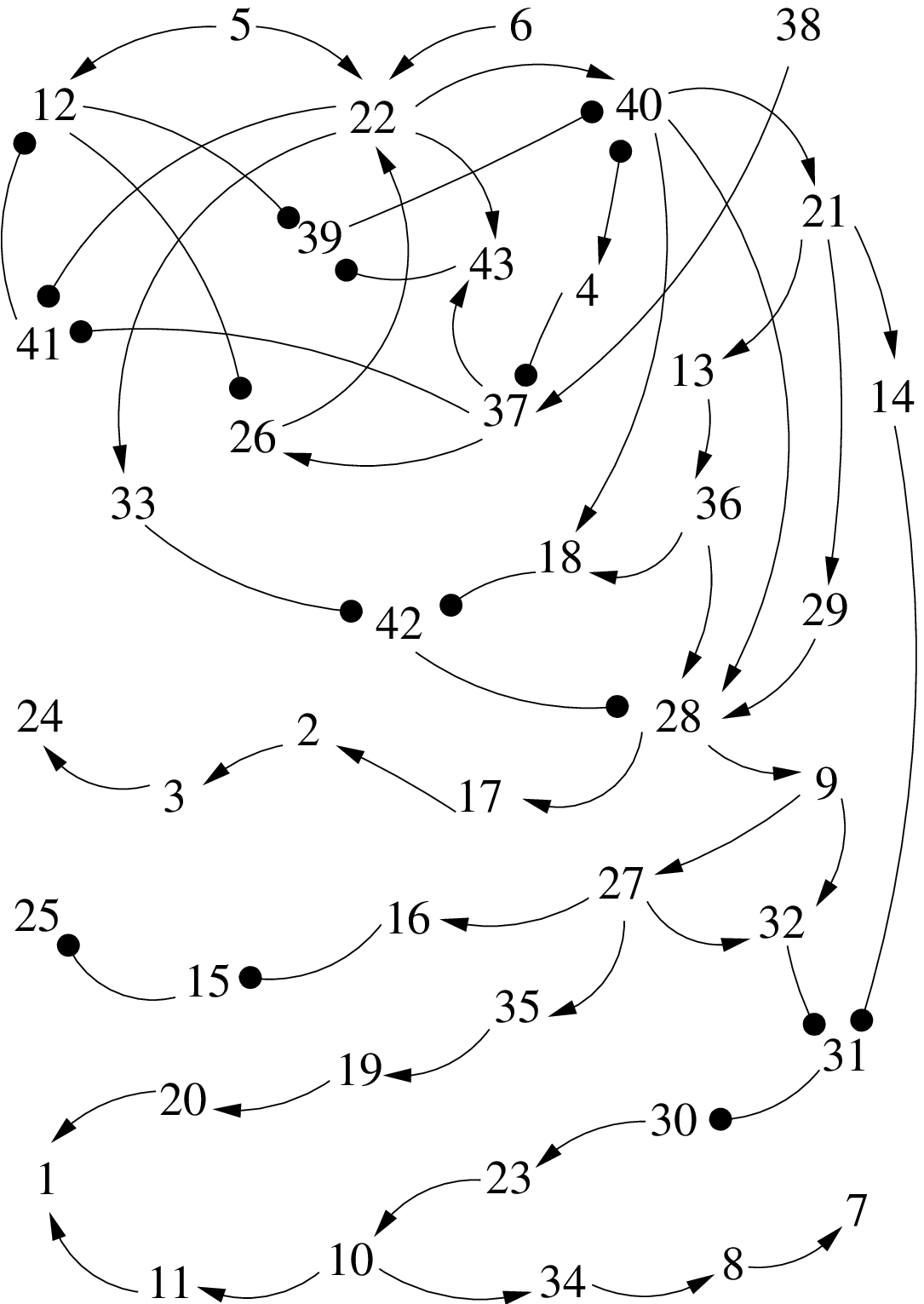}  
} 
\framebox{  
\includegraphics[totalheight=.4cm]{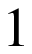}  
}}
\caption{AND-NOT model of T-cell receptor (left) and the reduced AND-NOT network (right). } \label{fig:tcellrecep}
\end{figure}

%%%%%%%%%%%%%%%%%%
Here we consider an AND-NOT network model of the T-cell receptor based on the Boolean model in \cite{tcellrecep} (left).  
The wiring diagram is shown in Figure \ref{fig:tcellrecep}. The state space of this model has $2^{43}\approx 8.8\times 10^{12}$ states.

$\begin{array}{lllllll}
f_{1} & = &  x_{11} \wedge  x_{20}  &, &
f_{2} & = &  x_{17 } \\
f_{3} & = &  x_{2 }  &, &
f_{4} & = &  x_{40 } \\
f_{5} & = & 1   &, &
f_{6} & = & 1  \\
f_{7} & = &  x_{8 }  &, &
f_{8} & = &  x_{34 } \\
f_{9} & = &  x_{28 }  &, &
f_{10} & = &  x_{23 } \\
f_{11} & = &  x_{10 }  &, &
f_{12} & = & \neg x_{41} \wedge  x_{5 } \\
f_{13} & = &  x_{21 }  &, &
f_{14} & = &  x_{21 } \\
f_{15} & = &  \neg x_{16 }  &, &
f_{16} & = &  x_{27 } \\
f_{17} & = &  x_{28 }  &, &
f_{18} & = &  x_{36} \wedge  x_{40} \\
f_{19} & = &   x_{35 }  &, &
f_{20} & = &  x_{19 } \\
f_{21} & = &   x_{40 }  &, &
f_{22} & = &   x_{26} \wedge  x_{5} \wedge  x_{6} \\ 
f_{23} & = &  x_{30 }  &, &
f_{24} & = &  x_{3 } \\
f_{25} & = &  \neg x_{15 }  &, &
f_{26} & = &  x_{37 } \wedge  \neg x_{12} \\
f_{27} & = &   x_{9 }  &, &
f_{28} & = &   x_{29} \wedge \neg x_{42} \wedge  x_{36} \wedge  x_{40 } \\
f_{29} & = &  x_{21 }  &, &
f_{30} & = & \neg x_{31 } \\
f_{31} & = & \neg x_{14 } \wedge  \neg x_{32 }  &, &
f_{32} & = &  x_{27} \wedge  x_{9 } \\
f_{33} & = &   x_{22 }  &, &
f_{34} & = &  x_{10 } \\
f_{35} & = &  x_{27 }  &, &
f_{36} & = &  x_{13} \\
f_{37} & = &   x_{38} \wedge \neg x_{4}  &, &
f_{38} & = &  1 \\ 
f_{39} & = & \neg x_{12} \wedge \neg x_{43 }  &, &
f_{40} & = &   x_{22} \wedge \neg x_{39} \wedge \neg x_{4} \\
f_{41} & = & \neg x_{22} \wedge \neg x_{37}  &, &
f_{42} & = & \neg x_{18} \wedge \neg x_{33} \\
f_{43} & = &  x_{22} \wedge  x_{37} \\
\end{array}$

By using our algorithm we reduce this AND-NOT network to the AND-NOT network shown in  Figure \ref{fig:tcellrecep} (right), $h:\{0,1\}^{1}\rightarrow\{0,1\}^{1}$ given by $h(x_{1})=1$.  Its state space has only 2 states, which is about 13 orders of magnitude smaller than the original state space.
It is easy to see that this network has a unique steady state ($x_{1}=1$). 
Therefore, the original AND-NOT model has a unique steady state. The timing of the reduction is $.00272086s$  (average of 1000000 repetitions).

%%%%%%%%%%%%%%%%%
\subsection{Random AND-NOT networks}\label{section:random}
\begin{figure}[here]
\centerline{  \framebox{ 
\ 
\includegraphics[totalheight=7cm]{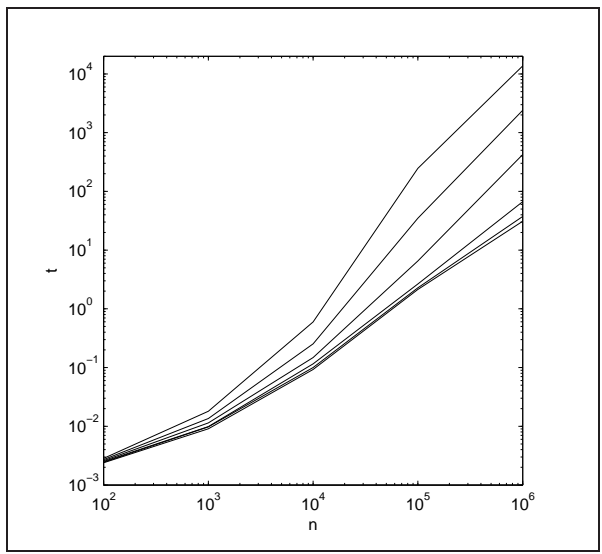}  \ 
} }
\caption{Average timing (in seconds) v.s. number of nodes for different values of $\gamma$. From top to bottom: $\gamma=2.0, 2.2, \dots, 3.0$.} \label{fig:time_vs_n}
\end{figure}

%%%%%%%%%%%%%%%%%%
In this section we show that our algorithm works very well for large sparse AND-NOT networks. We run our implementation of the algorithm on a Linux system using one 2.40GHz CPU core. To mimic wiring diagrams of gene regulatory networks, we considered random AND-NOT networks with wiring diagrams where the in-degree followed a power law distribution \cite{Huynen01051998, Aldana200345,Albert01112005} with no constant nodes. Since the parameter $\gamma$ in the power law distribution is usually between 2 and 3 for biochemical networks \cite{Aldana200345, Albert01112005}, we considered the parameters $\gamma=2.0, 2.2, 2.4, 2.6, 2.8, 3.0$. We analyzed about 100000 AND-NOT networks. The summary of the analysis for $\gamma=2$ and $\gamma=3$ is shown Table \ref{table:random}. Figure \ref{fig:time_vs_n} shows the plots of time ($t$) v.s. the size of the network ($n$) for $\gamma=2.0, 2.2, 2.4, 2.6, 2.8, 3.0$ in a log-log scale. These timings include the timing of the reduction steps and the timing of steady state computation, although the latter turned out to be negligible. More precisely, the number of nodes of the reduced AND-NOT networks, $m$, were very small with an average of $\mu_m=2.6$, ranging from $m=0$ to $m=19$. Since these numbers are small, exhaustive search was more than enough to compute the steady states of the reduced networks. It is important to mention that if the reduced AND-NOT network is too large to handle by exhaustive search, one can use additional tools such as polynomial algebra \cite{Veliz_poly}, but as mentioned before, this was not necessary for our simulations.

\begin{table}[here]
\caption{Timimg, $t$, for the reduction algorithm (in seconds). The mean and standard deviation of $t$ are denoted by $\mu_t$ and $\sigma_t$, respectively. Last row: best fit polynomial $t=cn^k$.}\label{table:random}
\begin{tabular}{|c|cc|cc|}
  \hline
      &  \multicolumn{2}{c|}{ $\gamma=2.0$} &  \multicolumn{2}{c|}{$\gamma=3.0$}  \\
     $n$  & $\mu_t$ & $\sigma_t$ &  $\mu_t$ & $\sigma_t$  \\   
  \hline
   $10^2$ & $.002838$ & $.000603$  & $.002398$ & $.000710$\\
  \hline
   $10^3$ & $.018034$ & $.003995$   &    $.009058$ & $.001744$  \\
  \hline
   $10^4$ & $.597764$ & $ .242391$   &    $.092602$ & $.015238$  \\
  \hline
   $10^5$ & $246.241$ & $147.727$   &    $2.13928$ & $.493422$  \\
  \hline
   $10^6$ & $13661.1$ & $9129.92$  &    $31.0311$ & $5.36648$  \\
   \hline
  Best fit of $t=cn^k$ & \multicolumn{2}{c|}{$c\approx 10^{-8}, k\approx 2.025$} & \multicolumn{2}{c|}{$c\approx 2\times 10^{-6}, k\approx 1.197$}  \\
  %c=1.0617(10^-8),k=2.0253, c=1.9991(10^-6), k=1.1968
  \hline
\end{tabular}
\end{table}

We can see in Figure \ref{fig:time_vs_n} that our reduction algorithm scales well with the number of nodes. Furthermore, our algorithm can reduce networks with 1000000 nodes. We also see that for very sparse networks (i.e. large values of $\gamma$) our algorithm scales very well. As sparsity is lost (i.e. as $\gamma$ decreases), our algorithm becomes less and less scalable; however, as mentioned before, the value of $\gamma$ for biochemical networks is usually between 2 and 3 for which our algorithm performs well. Also, the timings look polynomial (linear on the log-log scale), especially for large $\gamma$ and $n$. The best fit polynomial of the form $t=cn^k$ for $\gamma=2$ was given by $c\approx 10^{-8}$ and $k\approx 2.025$; and for $\gamma=3$ was given by $c\approx 2\times 10^{-6}$ and $k\approx 1.197$. Note: Although using the timings at $n=10^2$ for the estimation of $c$ and $k$ would give a smaller value of $k$, we did not use them because the linear relationship between $\log(t)$ and $\log(n)$ seemed to start at $n=10^3$.

%%%%%%%%%%%%%%%%%%%%%%%%%%%%%%%%%%%%%
\section{Discussion}
\label{sec-discussion}
%%%%%%%%%%%%%%%%%%%%%%%%%%%%%%%%%%%%%

Since the problem of analyzing BNs is hard for large networks, many reduction algorithms have been proposed \cite{redbn,redlm,Saadatpour2010641}. However, it is not clear if such algorithms scale well with the size of the network. In order to optimize reduction algorithms, it is necessary to focus on specific families of BNs. 

The family of AND-NOT networks has been proposed as a special family simple enough for theoretical analysis, but general enough for modeling \cite{bnandnot,SCBN,CBN}. Thus, we propose an  algorithm for network reduction for the family of AND-NOT networks. A key property of our algorithm is that it preserves steady states, so it can be very useful  in steady state analysis. We applied our algorithm to three AND-NOT network models, namely, Th-cell differentiation, ERBB2 activation, and T-cell receptor. Our reduction algorithm performed very well with these models; the state space of the reduced networks were several orders of magnitude smaller than the original state space. This greatly simplified steady state computation. Using random AND-NOT networks, we showed that our algorithm scales well with the number of nodes and can handle large sparse AND-NOT networks with up to 1000000 nodes. To the best of our knowledge, no other algorithm can handle AND-NOT networks or any other class of (nonlinear) BNs of this size.

It is important to mention that since our reduction algorithm is defined using the wiring diagram only, it has the special property that it runs in polynomial time. That is, we have developed a polynomial-time algorithm that reduces the problem of finding steady states of an AND-NOT network, $f:\{0,1\}^n\rightarrow \{0,1\}^n$, into the problem of finding the steady states of a smaller AND-NOT network, $g:\{0,1\}^m\rightarrow \{0,1\}^m$, where $m\leq n$. Also, the steady states of the reduced AND-NOT network can be used to compute the steady states of the original AND-NOT network in polynomial time. Thus, our algorithm transforms an NP-complete problem of input size $n$ into an NP-complete problem of input size $m$, where $m\leq n$. While this represents no theoretical improvement in the complexity of the problem, it does represent a significant improvement in the practical ability to analyze actual models that arise in molecular systems biology because they are very sparse, and in that case we typically have $m<<n$. Also, this could provide a novel way to solve NP-complete problems by first using our polynomial-time algorithm as a pre-processing step.

\section*{Appendix: Storing and using AND-NOT networks.}
We store AND-NOT networks as a text file using the following format.
{ \begin{verbatim}
num_nodes
num_edges
edge1
edge2
...
ZERO_NODES
zeronode1
zeronode2
...
\end{verbatim} }

In the file above \verb!num_nodes! is the number of nodes, \verb!num_edges! is the number of edges, \verb!edgei! is an edge written in the format ``\verb!input output sign!''. The nodes that have the Boolean function 0 are given below \verb!ZERO_NODES!. The next example shows this in more detail.

\textbf{Example from Section \ref{sec-red-imp}. }
The following is the file that stores the AND-NOT network.
{ 
\begin{verbatim}
6
8
4 1 1
1 2 1
3 2 -1
4 2 1
1 4 1
2 5 -1
4 5 1
6 5 1
ZERO_NODES
3
\end{verbatim}}

We feed this file to the reduction part of the algorithm and obtain the file \verb!reduced.txt!.
{
\begin{verbatim}
1
1
4  4  1
ZERO_NODES
3
ACYCLIC_GRAPH
2 5 -1
4 5 1
6 5 1
4 1 1
4 2 1
\end{verbatim}}

As before, the first line is the number of nodes, the second line is the number of edges (not counting edges from 0), and the numbers below \verb!ZERO_NODES! are the nodes that have the Boolean function 0. The reduced AND-NOT network is given by the first 5 lines, the rest of the file is the acyclic graph. 
We feed the file \verb!reduced.txt! to the steady state computation part of our algorithm and obtain the file \verb!ss_reduced.txt!. 
{ \begin{verbatim}
0
1
\end{verbatim}}

This means that there are two steady states, $x_4=0$ and $x_4=1$. We now feed \verb!ss_reduced.txt! and \verb!reduced.txt! to the backwards substitution part of our algorithm and obtain the file \verb!ss.txt!, which contains the steady states of the original AND-NOT network.
{ \begin{verbatim}
000001
110101
\end{verbatim}}

\textbf{Example from Section \ref{sec-application}.1. }
Here we show the input and output of our algorithm. The AND-NOT network is encoded as the file \verb!example1.txt! given below.

{ \begin{verbatim}
26
38        
22 1  -1
26 1  -1
2  3  -1
19 4  -1
24 4  -1
4  5   1
1  6   1
6  7   1
8  9  -1
21 9  -1
10 11 -1
21 11 -1
1  12  1
18 12  1
12 13  1
17 13  1
11 14  1
5  15  1
17 15  1
23 16 -1
22 17 -1
18 17  1
3  18 -1
15 18 -1
7  19  1
9  20  1
1  20 -1
13 21  1
1  22 -1
25 22 -1
14 24 -1
16 24 -1
20 24 -1
22 24 -1
22 25 -1
18 25  1
1  26 -1
21 26 -1
ZERO_NODES
\end{verbatim}}

By using our code (called \verb!AND_NOT_analysis!) we obtain the following.
{ \begin{verbatim}
user@comp:~$  AND_NOT_analysis < example1.txt
01000001010000001100001111
01011001010000000100011001
11000111010110001110101110
\end{verbatim}}

\end{document}